\documentclass[11pt,a4paper]{article}
\usepackage{amsmath,amssymb,latexsym,amscd,bbold}
\usepackage{jheppub, slashed, color,subfig}
\usepackage{enumitem}
\usepackage{multirow}
\usepackage{textcomp}
\usepackage{array} 
\usepackage{graphicx}
\usepackage{adjustbox}
\usepackage{mathrsfs}
\usepackage{dsfont}
\usepackage{parskip}

\newcommand{\R}{\mathbb{R}}

\newcommand{\del}{\partial}

\newcommand{\ra}{\rightarrow}
\newcommand{\Ra}{\Rightarrow}

\newcommand{\txd}{\textrm{d}}

\newcommand{\Tr}{~\textrm{Tr}}

\def\ie{\begin{equation}\begin{aligned}}
\def\fe{\end{aligned}\end{equation}}
\def\be{\begin{equation}\begin{aligned}}
\def\ee{\end{aligned} \end{equation}}

\renewcommand{\lg}{\mathfrak{g}}
\newcommand{\ltg}{\mathfrak{tg}}
\newcommand{\lh}{\mathfrak{h}}

\begin{document}

\title{Integrable Deformations of the Breitenlohner--Maison Model from 4d Chern--Simons Theory }
\author{Meer Ashwinkumar, Matthias Blau}
\affiliation{Albert Einstein Center for Fundamental Physics, Institute for Theoretical Physics,
University of Bern, Sidlerstrasse 5, CH-3012 Bern, Switzerland}
\emailAdd{meer.ashwinkumar@unibe.ch} \emailAdd{blau@itp.unibe.ch}
\abstract{We derive integrable deformations of the 2d
Breitenlohner--Maison (BM) sigma model that describes the stationary,
axisymmetric sector of 4d general relativity, as well as higher-rank
generalisations thereof, using
the framework of 4d Chern--Simons theory. In particular,  we consider deformations of the boundary conditions and action of the
4d Cole--Weck model, 
which lead to deformations of the BM model
associated with solutions to the homogeneous
and inhomogeneous classical Yang--Baxter equations respectively.}

\maketitle

\section{Introduction}

The study of black holes in theories of gravity is facilitated by
the integrability of such theories with $D-2$ commuting Killing
symmetries. These Killing symmetries allow for powerful exact
solution generating techniques, since they imply that the gravitational
dynamics reduce to that of a two-dimensional nonlinear sigma model.
This nonlinear sigma model is integrable in the sense that it admits
a flat Lax connection.  For example, the stationary, axisymmetric
sector of four-dimensional general relativity is well-known to be
described by a two-dimensional integrable sigma model with target
space $\textrm{SL}(2,\mathbb{R})/\textrm{SO}(2)$ coupled to 2d
gravity, also known as the Breitenlohner--Maison or BM model
\cite{Breitenlohner:1986um}, while the reduction of $D$-dimensional
gravity leads to an analogous sigma-model with target space
$\textrm{SL}(D-2,\mathbb{R})/\textrm{SO}(D-2)$ \cite{Breitenlohner:1987dg}.
The special feature of these (generalised) BM models, that sets it apart
from the standard symmetric space sigma model actions, is a characteristic 
world-sheet dependence of the coupling parameter.

The existence of a Lax connection was used by Belinski and Zakharov to generate 4d vacuum solutions to Einstein's equations describing multiple black holes with a common axis of symmetry \cite{Belinsky:1971nt, Belinsky:1979mh}. The generalisation of their techniques to higher dimensions has moreover led to interesting solutions such as black rings and black Saturns \cite{Pomeransky:2005sj}. 

More recently, a unifying framework for integrable systems has emerged  in the
form of a 4d variant of Chern--Simons (CS) theory \cite{Costello:2013zra, Costello:2017dso, Costello:2018gyb, Costello:2019tri}. In particular, the work of Costello and Yamazaki showed that this 4d Chern--Simons theory could be used as an efficient unifying framework for integrable field theories. In 4d Chern--Simons theory, the spectral parameter that enters the Lax connection is identified with a point on a Riemann surface that is a submanifold of the background on which the theory is defined.

Using the techniques of Costello and Yamazaki \cite{Costello:2019tri}, Cole and
Weck derived the sigma model sector of the BM model (as well its higher rank
generalisations for other Lie algebras $\lg$) from 4d Chern--Simons (CS)
theory \cite{Cole:2024skp}, and further elucidated its origin from twistor
space. 
Given the computational power furnished by 4d Chern--Simons theory,
it stands to reason that integrable deformations of the BM model
should also be derivable from 4d Chern--Simons theory. In the following, we
will look at deformations of this sigma model sector of the BM model. 

Some examples of such integrable deformations have appeared in the work of
Ces{\`a}ro and Osten \cite{Cesaro:2025msv}, including a proposal for
gravitational analogues of the homogeneous and inhomogeneous Yang--Baxter
deformation of the principal chiral model studied in \cite{Klimcik:2002zj,
Klimcik:2008eq, Matsumoto:2015jja}. Deformations of theories of gravity with
$D-2$ commuting Killing symmetries, such as by the addition of a cosmological
constant or a dilaton potential, often break integrability of the effective
two-dimensional sigma model. The study of integrable deformations of the BM
model thus constitutes an important endeavour, given that such deformations ought to correspond to interesting deformations of theories of gravity.


Our aim in this work is to derive the homogeneous and inhomogenous
Yang--Baxter deformation of the BM model studied in \cite{Cesaro:2025msv}
from 4d Chern--Simons theory. In particular, we will show that the
former are related to mixed Dirichlet-Neumann deformations of the
original Dirichlet boundary condition employed by Cole and Weck,
while the latter arise from a suitable deformation of the 4d
Chern--Simons action itself (via a deformation of the meromorphic
1-form $\omega$ that appears in the 4d Chern--Simons action). For
gauge group $G$ and corresponding Lie algebra $\lg$, the algebraic
structure associated with each boundary condition in the homogeneous
case turns out to be a worldsheet-dependent version of a subalgebra
of the Lie algebra $\ltg$ of the tangent bundle group $TG$, that is
associated with a solution of the classical Yang--Baxter equation.
In the inhomogenous case, the algebraic structure associated with
each boundary condition is a subalgebra of $\lg \oplus \lg$ associated
with a solution of the modified classical Yang--Baxter equation.

The results presented in this work may be viewed as evidence that 4d
Chern--Simons theory provides a systematic framework for deriving deformations
of the BM model. We expect that the methods developed in this work will lead to
the construction of many further such deformations, such as
analogues of the $\lambda$-model and E-model \cite{Klimcik:2015gba}. Moreover, since the BM model itself arises from a symmetry reduction of a 4d integrable field theory, which in turn can be obtained from 6d Chern--Simons theory on twistor space \cite{Cole:2024skp}, it is natural to expect that its integrable deformations admit a similar higher-dimensional origin. 4d Chern--Simons theory 
could also be used to investigate aspects of quantisation of the BM model and its deformations. 
It would also be very interesting to find a gravitational interpretation for the
deformed BM models we have derived.

We emphasise that the 4d Chern--Simons approach
to deriving deformations of the BM model has multiple advantages.
Firstly, the derivation of such models from 4d Chern--Simons theory
furnishes their corresponding Lax operators as classical gauge
fields. The Lax operators would otherwise require educated guesswork
to be obtained. The embedding of deformations of the BM model into
4d Chern--Simons theory is expected to facilitate derivations of
their generalisations, especially those involving multiple fields.
In addition, 4d Chern--Simons theory provides a framework for the
derivation of dualities of these theories, such as bosonisation and
Poisson--Lie T-duality \cite{Delduc:2019whp,Ashwinkumar:2023zbu}.
The 4d Chern--Simons approach also elucidates the underlying algebraic
structures of the deformations of the BM model that arise as part
of the 4d boundary conditions.  The embedding of 4d Chern--Simons
theory in 6d twistor space moreover elucidates the geometric origin
of these models and their relationship with 4d integrable field
theories. 

The rest of this paper is organised as follows. In Section \ref{Sec2},
we review the 4d CS action studied used by Cole and Weck to derive
the BM model. In Section \ref{Sec3}, we proceed to derive the
deformation of the BM model associated with solutions to the
inhomogeneous modified classical Yang--Baxter equation. In Section
\ref{Sec4}, we derive the deformation of the BM model associated
with the homogeneous classical Yang--Baxter equation. Some technical details are
discussed in Appendices \ref{derwz} - \ref{subscd}.

\textbf{Acknowledgement}: We would like to thank David Osten for a useful discussion. The research of M.A.\ is supported in part by the NCCR SwissMAP of the Swiss National Science Foundation.

\section{Review of the 4d CS Action of Cole and Weck}
\label{Sec2}

The starting point of the work of Cole and Weck \cite{Cole:2024skp} is
the 4d CS action 
\ie \label{1staction}
S_{\omega CS}[A]=\frac{1}{2 \pi \mathrm{i}} \int_{\mathbb{R}^2\times\mathbb{C P}^1 } \omega \wedge \operatorname{Tr}\left(A \wedge \mathrm{d} A+\frac{2}{3} A \wedge A \wedge A\right)
\fe 
with
\ie \label{omeggcw}
\omega =  \left(\frac{\mathrm{i}(\xi-\bar{\xi})}{4}\left(\frac{Z^2+1}{Z^2} \mathrm{~d} Z\right)+\frac{\mathrm{i}(Z-\mathrm{i})^2}{4 Z} \mathrm{~d} \xi-\frac{\mathrm{i}(Z+\mathrm{i})^2}{4 Z} \mathrm{~d} \bar{\xi}\right ),
\fe 
where $\xi =z+i\rho $ and $\bar{\xi}=z-i\rho$ are complex worldsheet coordinates
on $\mathbb{R}^2$ (following Cole and Weck we are working in Euclidean 
signature) and $Z$ is a holomorphic local coordinate on
$\mathbb{CP}^1$. The other data entering the Cole--Weck model are a $G$ gauge field $A$
for some choice of gauge group $G$ with semi-simple 
Lie algebra $\lg$ (with invariant scalar product denoted by $\operatorname{Tr}$), 
and a $\mathbb{Z}_2$-automorphism $\eta$ of $\lg$ (whose significance will be explained
below).

The main difference between the Cole--Weck model
and  4d Chern--Simons constructions of integrable field theories
that appear in \cite{Costello:2019tri} are the spacetime $\xi$ and $\bar{\xi}$
components of $\omega$. As explained by Cole and Weck
\cite{Cole:2024skp}, this novel feature of the 1-form $\omega$ 
is required in order to reproduce the two characteristic features of 
the BM model, namely the spacetime dependence of the coupling constant, and 
the mixing of the spectral parameter and the spacetime coordinates in the 
Lax connection.

We can equivalently write the meromorphic 1-form as  
$\omega = \textrm{d}W$, where  
\ie \label{wdef}
W=z+\frac{\rho}{2}\left(Z^{-1}-Z\right),
\fe
with a branch cut between $W=z+i\rho $ and $W=z-i\rho $. These branch points correspond to $Z=-i$ and $Z=i$ respectively, which are critical points of the map defined by \eqref{wdef}, that is, $\frac{\txd W}{\txd Z}=0$ at these points. Notice that for $Z\rightarrow -\frac{1}{Z}$, we find $W\rightarrow W$. The $Z$-coordinate thus parametrises a two-sheeted covering of a Riemann surface locally parametrised by $W$ and is given explicitly by 
\ie Z=\frac{1}{\rho}\left(z-W \pm \sqrt{(W-z)^2+\rho^2}\right).
\fe 
According to the Riemann-Hurwitz theorem, the Riemann surface parametrised by $W$ also has genus 0, i.e., it can be identified with $\mathbb{CP}^1$.

The gauge field is restricted such that it can be thought of as a
single-valued gauge field on the $W$-plane (the local patch on
$\mathbb{CP}^1$ parametrised by $W$), while being constrained on
the $Z$-plane. The constraint is of the form 
\ie \label{equi}
A(Z)=\eta A(-1/Z),
\fe 
where $\eta : \mathfrak{g} \rightarrow
\mathfrak{g}$ is a $\mathbb{Z}_2$-automorphism.
In the specific case of $\mathfrak{g}=\mathfrak{sl}(2,\mathbb{R})$
relevant to the stationary and axisymmetric sector of general
relativity, and hence to the original BM model, $\eta$ is defined
as $x \mapsto-x^T$, and the $\eta$-invariant subalgebra of $\lg$ 
is $\mathfrak{so}(2)$, eventually leading to the BM
$\mathrm{SL}(2,\R)/\mathrm{SO}(2)$ target space.
We shall mainly
focus on the description of the 4d CS theory on the $Z$-plane in
what follows, and only use the $W$-plane description where it is
convenient.

The action \eqref{1staction} can be rewritten as \ie \label{2ndaction}
S_{\omega CS}[A]&= \frac{1}{2 \pi \mathrm{i}}
\int_{\mathbb{R}^2\times\mathbb{C P}^1 } \omega_Z\textrm{d}Z \wedge
\operatorname{Tr}\left(A \wedge {{D}} A+\frac{2}{3} A \wedge A
\wedge A\right)\\ &+\frac{1}{2 \pi
\mathrm{i}}\int_{\mathbb{R}^2\times\mathbb{C P}^1 }\textrm{d}^4x
\textrm{Tr}\left((2A_Z)(\omega_{\xi}F_{\bar{Z}\bar{\xi}}-
\omega_{\bar{\xi}}F_{\bar{Z}{\xi}})\right)\\ &+\frac{1}{2 \pi
\mathrm{i}}\int_{\mathbb{R}^2\times\mathbb{C P}^1 }\textrm{d}^4x
\textrm{Tr}\left(  A_Z
(-\partial_{\bar{Z}}\omega_{\xi}A_{\bar{\xi}}+\partial_{\bar{Z}}\omega_{\bar{\xi}}A_{{\xi}})\right)
, \fe where $\textrm{d}^4x=\textrm{d}Z\wedge \textrm{d}\bar{Z}\wedge
\textrm{d}\xi\wedge \textrm{d}\bar{\xi}$.  Here, we have utilised
the differential \ie \label{DDiff}{D}\equiv\textrm{d}\xi
\tilde{\partial}_{\xi}+ \textrm{d}\bar{\xi} \tilde{\partial}_{\bar{\xi}}
+ \textrm{d}Z {\partial}_{Z}+ \textrm{d}\bar{Z} {\partial}_{\bar{Z}},\fe
which involves the derivatives \begin{equation} \begin{aligned}
\label{shiftedders}
 \tilde{\partial}_{{\xi}}\equiv
 &\partial_{{\xi}}-\frac{\omega_{{\xi}}}{\omega_Z} \partial_Z
=\partial_{\xi}+(\xi-\bar{\xi})^{-1} \frac{1+\mathrm{i} Z}{1-\mathrm{i}
Z} Z \partial_Z \\
 \tilde{\partial}_{\bar{\xi}} \equiv
 &\partial_{\bar{\xi}}-\frac{\omega_{\bar{\xi}}}{\omega_Z}
 \partial_Z=\partial_{\bar{\xi}}-(\xi-\bar{\xi})^{-1} \frac{1-\mathrm{i}
 Z}{1+\mathrm{i} Z} Z \partial_Z,
\end{aligned} \end{equation} where $\omega_Z$, $\omega_{\xi}$ and
$\omega_{\bar{\xi}}$ are components of $\omega$ given in \eqref{omeggcw}.
As explained in Appendix \ref{derwz}, we have
\ie\tilde{\partial}_{{\xi},{\bar{\xi}}}={\partial}_{{\xi},{\bar{\xi}}}|_W\fe
and \ie{\partial}_{{\xi},{\bar{\xi}}}={\partial}_{{\xi},{\bar{\xi}}}|_Z,\fe
where $|_W$ and $|_Z$ indicate the variable being held fixed.

The third term in \eqref{2ndaction} localises to boundary terms at $Z=0$ since these are locations of simple poles of $\omega_{\xi}$ and $\omega_{\bar{\xi}}$.
Hence, performing a variation of the gauge field, we obtain 
the bulk equations of motion 
\begin{equation}
\begin{aligned}
\label{4eom}
&\omega_Z\left(\tilde{\partial}_{\xi} A_{\bar{Z}}-\partial_{\bar{Z}} A_{\xi}+\left[A_{\xi}, A_{\bar{Z}}\right]\right) =0\\
& \omega_Z\left(\tilde{\partial}_{\bar{\xi}} A_{\bar{Z}}-\partial_{\bar{Z}} A_{\bar{\xi}}+\left[A_{\bar{\xi}}, A_{\bar{Z}}\right]\right)=0 \\
& \omega_Z\left(\tilde{\partial}_{{\xi}} A_{\bar{\xi}}-\tilde{\partial}_{\bar{\xi}} A_{\xi}+\left[A_{\xi}, A_{\bar{\xi}}\right]\right)=0 \\
& -\omega_{\xi}F_{\bar{Z}\bar{\xi}}+\omega_{\bar{\xi}}F_{\bar{Z}{\xi}}=0.
\end{aligned}
\end{equation}
The final equation 
in \eqref{4eom} arises from the variation of $A_Z$, and can be derived from the first two equations in \eqref{4eom}, meaning that there are only three independent equations of motion. 
This reflects the fact that there is a gauge symmetry that can be used to set $A_Z=0$.
Indeed, the 4d CS action is invariant under the shift 
\ie 
A\rightarrow A+ C \omega\fe 
for some constant $C$.
Since $\omega=\textrm{d}W$, this allows us to set $A_W=0$.
Although the $Z$-plane is a double cover of the $W$-plane, on a given local branch we can perform the change of coordinates from $(W,\xi,\bar{\xi})$ to $(Z,\xi,\bar{\xi})$, and we find that $A_Z$ is related to $A_W$ as  \begin{equation}
A_Z=A_W \frac{\partial W}{\partial Z}+A_{\bar{W}} \frac{\partial \bar{W}}{\partial Z}+A_{\xi} \frac{\partial \xi}{\partial Z}+A_{\bar{\xi}} \frac{\partial \bar{\xi}}{\partial Z} = A_W \frac{\partial W}{\partial Z}.
\end{equation}
This means that $A_W=0$ implies $A_Z=0$.

The action \eqref{2ndaction} simplifies in the gauge $A_Z=0$ to
\ie \label{2ndactionB}
S_{\omega CS}[A] \ra 
\frac{1}{2 \pi \mathrm{i}} \int_{\mathbb{R}^2\times\mathbb{C P}^1 } \omega_Z\textrm{d}Z \wedge \operatorname{Tr}\left(A \wedge {{D}} A+\frac{2}{3} A \wedge A \wedge A\right),
\fe 
which takes the form of a 4d Chern--Simons action with meromorphic 1-form $\omega_Z\textrm{d}Z$, and with the exterior derivative $\textrm{d}$ replaced by $D$ defined in \eqref{DDiff}. This form of the action is convenient for derivations of integrable field theories, and shall be used as the starting point in subsequent sections.

The derivation of an integrable field theory from the action \eqref{2ndactionB} involves 
setting 
\ie 
A_{\bar{Z}}=\hat{g}^{-1}\partial_{\bar{Z}}\hat{g}
\fe 
and performing 
the change of variables (often referred to as a formal gauge transformation, that is, a gauge transformation that does not preserve the boundary conditions)
\begin{equation}\label{form}
A_i\equiv 
\hat{g}^{-1} \mathcal{L} \hat{g}+\hat{g}^{-1} \tilde{\partial}_i \hat{g},
\end{equation}
where $i=\xi,\bar{\xi},\bar{Z}$ and $\mathcal{L}_{\bar{Z}}=0$.
Under this change of variables, the equations of motion take the form
\begin{equation}
\begin{aligned}
\label{Leqs}
&\omega_Z\partial_{\bar{Z}} \mathcal{L}_{\xi} =0\\
& \omega_Z\partial_{\bar{Z}} \mathcal{L}_{\bar{\xi}}=0 \\
& \omega_Z\left(\tilde{\partial}_{{\xi}} \mathcal{L}_{\bar{\xi}}-\tilde{\partial}_{\bar{\xi}} \mathcal{L}_{\xi}+\left[\mathcal{L}_{\xi}, \mathcal{L}_{\bar{\xi}}\right]\right)=0.
\end{aligned}
\end{equation}
Here, $\mathcal{L}$ can be identified with the Lax connection of the integrable field theory.

The action \eqref{2ndaction} is also invariant under a gauge transformation that leaves the boundary conditions and the condition $A_Z=0$ invariant, where the gauge transformed connection is 
\ie \label{gaugetxtild}
A_i^U= U^{-1}A_iU+ U^{-1}\tilde{\partial}_iU
\fe 
for $i=\xi,\bar{\xi},\bar{Z}$ and $U:\Sigma \times C\rightarrow G$.
In particular, such a gauge transformation can be understood as transforming $\hat{g}$ that appears in the change of variables \eqref{form} as 
\ie \label{rigmul}
\hat{g} \rightarrow \hat{g}U.
\fe 
The transformation \eqref{rigmul} is useful in the derivation of IFTs to remove redundant degrees of freedom that arise due to gauge invariance of the boundary conditions at poles of $\omega_Z \textrm{d}Z$.


Following \cite{Delduc:2019whp}, the formal gauge transformation of \eqref{2ndaction} leads to a unifying 2d action of the form 
\ie \label{grauni}
\sum_{x \in poles} \int_{\mathbb{R}^2}\textrm{Tr}\left(\operatorname{res}_x \omega_Z\textrm{d}Z \wedge \mathcal{L}\wedge \tilde{\textrm{d}} g_x g_x^{-1} \right) -\sum_{x \in poles}\left(\operatorname{res}_x \omega_Z\textrm{d}Z\right) I_{\mathrm{WZ}}\left[g_x\right]
\fe 
where the sums are over the poles of $\omega_Z\textrm{d}Z$, and $g_x$ denotes the restriction of $\hat{g}$ to the pole labelled $x$, and where $I_{WZ}[g_x]$ denotes the Wess--Zumino term that is given explictly by 
\ie I_{\mathrm{WZ}}\left[g_x\right]:=-\frac{1}{3} \int_{{\mathbb{R}^2} \times\left[0, 1\right]}\textrm{Tr}\left(\hat{g}_x^{-1} d \hat{g}_x\wedge  \hat{g}_x^{-1} d \hat{g}_x \wedge \hat{g}_x^{-1} d \hat{g}_x\right).\fe 
Since the Lax operator typically involves the Maurer-Cartan current, this unifying action resembles a 2d Wess--Zumino--Witten action. 

Cole and Weck \cite{Cole:2024skp} impose Dirichlet boundary conditions on the
gauge field at $Z=0$ and $Z=\infty$. The equivariance condition \eqref{equi} 
then restricts the target space of the sigma model from $G$ to $G/H$, where 
the Lie algebra $\lh$ of $H$ is the $\eta$-invariant subalgebra of $\lg$, 
\ie
\label{hdef}
\lh = \{ x \in \lg:\; \eta(x) = x\}
\fe
In this way, the original BM model
can be derived from \eqref{grauni}
for $\lg =\mathfrak{sl}(2,\R)$. It has the action 
\ie \label{bmmodel}
\int_{\mathbb{R}^2} \mathrm{d} \xi \wedge \mathrm{d} \bar{\xi}(-i\rho ) \operatorname{Tr}\left((g^{-1}\partial_{\xi} g -\tilde{g}^{-1}\partial_{\xi} \tilde{g} )(g^{-1}\partial_{\bar{\xi}} g -\tilde{g}^{-1}\partial_{\bar{\xi}} \tilde{g} )\right)
\fe 
where $g$ and $\tilde{g}$ are edge modes associated with the poles of $\omega_Z
\textrm{d}Z$. This action can be entirely reexpressed in terms of the
$\mathrm{SL}(2, \mathbb{R}) / \mathrm{SO}(2)$ field $G=\tilde{g}^{-1}g$.
Moreover, the third equation of \eqref{Leqs} takes the form of the
familiar flatness condition of the Lax connection of the BM model
which involves the derivatives $\tilde{\partial}_{\xi}$ and
$\tilde{\partial}_{\bar{\xi}}$ defined in \eqref{shiftedders}.

More generally, for any $\lg$ and $\eta$, one obtains a BM-type sigma model
with symmetric target space $G/H$, and we will refer to all of these models
as BM models in the following.

\section{Gravitational Inhomogeneous Yang--Baxter Sigma Model}\label{Sec3}

In this section, we shall derive a deformation of the (sigma model sector of
the) BM model involving a solution of the inhomogeneous modified classical Yang--Baxter equation. 
 The inhomogeneous Yang--Baxter sigma model that deforms the principal chiral model
can be derived from 4d CS with boundary conditions involving a solution of the modified classical Yang--Baxter equation \cite{Delduc:2019whp}, and a deformation of the 1-form $\omega$ used to derive the principal chiral model. Hence, it is natural to expect that a deformation of the 1-form $\omega$ defined in \eqref{omeggcw} would lead us to an inhomogeneous Yang--Baxter deformation of the BM model.

We are thus led to consider the most general choice of $\omega$ on the $W$-plane. As explained in Appendix \ref{genomega}, by analysing the origin of the 4d CS considered here from twistor space by symmetry reductions with respect to Killing vector fields that generate translations and rotations, we can show that the most general choice of $\omega$ is
\ie 
\omega=F(W)\txd W,
\fe
where $F(W)$ is a function of $W$. Although a natural first choice for $F(W)$ is a polynomial in $W$, this would lead to poles of order three and higher in $Z$. However, most canonical examples of IFTs arise from poles of order one and two, and moreover,  4d CS with poles of order three or higher in $\omega$ requires a regularisation procedure to be well defined \cite{Benini:2020skc}.

We shall instead consider an $F(W)$ that gives us simple poles on the $Z$-plane, and this is the case of a single simple pole on the $W$-plane :
\ie \label{wsimplepole}
\omega = \frac{\textrm{d}W}{1-cW}
\fe 
(for some constant $c$, which for simplicity we choose to be real in the
following, $c \in \R$). 
Now, on the $Z$-plane, this $\omega$ is equivalent to 
\ie 
\omega =  \frac{1}{1-c(2z + \rho \left( \frac{1-Z^2}{Z}\right))}\left(\frac{\mathrm{i}(\xi-\bar{\xi})}{4}\left(\frac{Z^2+1}{Z^2} \mathrm{~d} Z\right)+\frac{\mathrm{i}(Z-\mathrm{i})^2}{4 Z} \mathrm{~d} \xi-\frac{\mathrm{i}(Z+\mathrm{i})^2}{4 Z} \mathrm{~d} \bar{\xi}\right ).
\fe 
Despite the deformation of $\omega$, the 4d CS action in the $A_Z=0$ gauge can still be presented in the form \eqref{2ndactionB}, since the deformation does not affect the shifted derivatives \eqref{shiftedders} because they depend on ratios of components of $\omega$.

We can thus proceed with the analysis using the 4d CS action of the form \eqref{2ndactionB}, where 
\ie \label{omzdzdef}
\omega_Z\textrm{d}Z  =  \frac{1}{1-c(2z + \rho \left( \frac{1-Z^2}{Z}\right))}\left(\frac{\mathrm{i}(\xi-\bar{\xi})}{4}\left(\frac{Z^2+1}{Z^2} \mathrm{~d} Z\right)\right ).
\fe 
Here, $\omega_Z\textrm{d}Z$ has zeroes at $\pm i$, and simple poles at
\ie 
\begin{aligned}
& Z = 0 \\
& Z =\frac{-1+2 c z-\sqrt{4 c^2 \rho^2+(1-2 c z)^2}}{2 c \rho} =:Z_+\\
& Z =\frac{-1+2 c z+\sqrt{4 c^2 \rho^2+(1-2 c z)^2}}{2 c \rho} =:Z_-.
\end{aligned}
\fe 
Since $\sqrt{4c^2 \rho^2 +(1-2cz)^2}= (1-2cz)\sqrt{\frac{4c^2\rho^2}{(1-2cz)^2}+1}$, we observe that in the $c\rightarrow 0 $ limit, $Z_-\rightarrow 0$ and $Z_+\rightarrow \infty$. 

Notably, the points $Z_+$ and $Z_-$ both map to $W=\frac{1}{c}$, which is the simple pole of \eqref{wsimplepole}. Also, $\infty$ and $0$ on the $Z$-plane map to $\infty$ on the $W$-plane. In other words, the four simple poles of $\omega$ on the $Z$-plane map to the two simple poles of \eqref{wsimplepole} on the $W$-plane, in line with the fact that the $Z$-plane is a double cover of the $W$-plane.

We can also make the following interesting observation. The meromorphic 1-form \eqref{omzdzdef} can be written as 
\ie \label{ommdz}
\omega_Z\textrm{d}Z=-\frac{1}{2} \frac{(Z^2+1) \textrm{d}Z}{c Z(Z-Z_+)(Z-Z_-)}.
\fe 
Let us evaluate the residues of $\omega_Z\textrm{d}Z$ at the four simple poles. We find 
\ie 
\textrm{res}_{Z=0} \omega_Z\textrm{d}Z &= 
\textrm{res}_{Z=\infty } \omega_Z\textrm{d}Z= \frac{1}{2 c }\\
\textrm{res}_{Z=Z_+}\omega_Z\textrm{d}Z &= 
\textrm{res}_{Z=Z_- } \omega_Z\textrm{d}Z= -\frac{1}{2 c }.
\fe 
In other words, there are two pairs of poles which have residues that are equal and opposite. 

The reason for this is as follows. The residue at $Z=0$ and $Z=Z_-$ should sum to zero since $Z_-$ approaches $0$ in the $c\rightarrow 0$ limit. The residues at $Z=\infty$ and $Z=Z_+$ should analogously sum to zero since $Z_+\rightarrow \infty$ in the $c\rightarrow 0$ limit. In addition, since $\omega_Z\textrm{d}Z$ has the same form close to $Z=0$ and $Z=\infty$, the residues at these points should be equal.

As in the derivation of the  ordinary inhomogeneous Yang--Baxter sigma model \cite{Delduc:2019whp}, we 
shall use boundary conditions involving solutions of the modified classical Yang--Baxter equation  to derive the inhomogeneous Yang-Baxter deformation of the BM model. 
Let us now specify these boundary conditions at the poles of $\omega_{Z}\textrm{d}Z $.
We shall choose, for antisymmetric matrices $R$ and $\tilde{R}$, the boundary conditions
\ie \label{ybbc}
(A_i|_{\infty},A_i|_{Z_+})&\in \mathfrak{g}_R \\
(A_i|_{0},A_i|_{Z_-})&\in \mathfrak{g}_{\tilde{R}} 
\fe 
where $i=\xi,\bar{\xi}$, and where
\ie \label{ralgss}
\mathfrak{g}_R&:=\{((R-1) x,(R+1) x) \mid {x} \in \mathfrak{g}\},\\
\mathfrak{g}_{\tilde{R}}&:=\{((\tilde{R}-1) \tilde{x},(\tilde{R}+1) \tilde{x}) \mid \tilde{x} \in \mathfrak{g}\}
\fe 
with $R$ and $\tilde{R}$   are related as 
\ie \label{310}
\tilde{R} \equiv \eta \circ R \circ \eta\; ,
\fe 
ensuring that the boundary conditions satisfy the equivariance condition
\eqref{equi}.
These boundary conditions satisfy the boundary equation of motion 
\begin{equation}
\sum_{x_+,x_-}\bigg(\left(\operatorname{res}_{x_{+}} (\omega_Z\textrm{d}Z)\right) \epsilon^{i j} \textrm{Tr}(A_i|_{x_{+}} \wedge \delta A_j|_{x_{+}} )+\left(\operatorname{res}_{x_{-}} (\omega_Z\textrm{d}Z)\right) \epsilon^{i j} \textrm{Tr}( A_i|_{x_{-}}\wedge \delta A_j|_{x_{-}} )\bigg)=0 
\end{equation}
for the pairs $(x_+,x_-)$ of poles where $\omega_Z\textrm{d}Z$ has equal and opposite residues, which follows from the fact that $R$ and $\tilde{R}$ are antisymmetric matrices.

In \eqref{ybbc}, we pair $0$ and $Z_-$ because $Z_-$ approaches 0 in the small $c$ limit. 
Note that when $Z_- \rightarrow 0$, the first boundary condition is equivalent to the Dirichlet boundary condition 
\ie 
A|_0=0,
\fe
and in the limit where $Z_+ \rightarrow \infty$, the second boundary condition becomes 
\ie 
A|_{\infty}=0,
\fe 
which are the boundary conditions used to derive the BM model in \cite{Cole:2024skp}.
The pair of boundary conditions \eqref{ybbc} can be identified with a single boundary condition on the $W$-plane, which is 
\ie 
(A_i|_{W=\infty},A_i|_{W=\frac{1}{c}})&\in \mathfrak{g}_R .
\fe

 Let us now further specify the data of the boundary conditions. We shall impose that $R$ and $\tilde{R}$ are  solutions of the inhomogeneous modified classical Yang-Baxter equation (mcYBE). For example, $R$ satisfies 
\ie \label{mcybe}
[R {x}, R {y}]-R([R {x}, {y}]+[{x}, R {y}])=-\tilde{c}^2[{x}, {y}]
\fe 
for ${x},{y} \in \mathfrak{g}$, where we shall choose $\tilde{c}$ to be 1. Noting that  \eqref{ybbc} are conditions on pairs of gauge fields, and each pair can be thought of as a gauge field for the group $G\times G$, restricting $R$ and $\tilde{R}$ to be solutions of \eqref{mcybe}
implies that the algebras \eqref{ralgss} are \textit{subalgebras} of $\mathfrak{g}\oplus \mathfrak{g}$ : 
\ie \label{rsolinhom}
& R\textrm{ solution of mcYBE}\quad \leftrightarrow  \quad \mathfrak{g}_R \subset \mathfrak{g}\oplus \mathfrak{g}|_{\infty, Z_+} , \, \\& \tilde{R}\textrm{ solution of mcYBE}\quad\leftrightarrow \quad \mathfrak{g}_{\tilde{R}} \subset \mathfrak{g}\oplus \mathfrak{g}|_{0, Z_-}.
\fe  
This further implies that the boundary conditions \eqref{ybbc} are gauge invariant under transformations of the form \eqref{gaugetxtild}
defined
with respect to the Lie groups corresponding to the Lie algebras \eqref{ralgss},
denoted $G_R$ and $G_{\tilde{R}}$ respectively. Note that since $R$ and ${\tilde{R}}$ are related as in \eqref{310} by a Lie algebra automorphism $\eta$, the algebras $\mathfrak{g}_R$ and  $\mathfrak{g}_{\tilde{R}}$ and the groups $G_R$ and  $G_{\tilde{R}}$ are isomorphic.

As briefly recalled in the previous section, a convenient method to derive an IFT from 4d CS is to first perform a formal gauge transformation (that does not preserve the boundary conditions) of the form \eqref{form}. Crucially, the boundary conditions constrain the equations of motion that are solved to obtain a Lax connection.
Now, since  $(G\times G)/G_R$ and $(G\times G)/G_{\tilde{R}}$ both correspond to the space of identical pairs in $G\times G$, the gauge symmetry of the boundary conditions can be fixed by identifying the edge modes as 
\ie \label{pairfix}
\hat{g}_{0,Z_-}=g, \quad \hat{g}_{\infty,Z_+}=\tilde{g}.\\
\fe
for some $g, \tilde{g}: \Sigma \rightarrow G$.
The gauge invariance at the simple poles however does not constrain the restrictions of the holomorphic derivatives of $\hat{g}$ to the poles. We shall choose to fix them to zero, as in the derivation of the BM model in \cite{Cole:2024skp}. 

Let us now proceed to derive an IFT. We shall first make a convenient change of variables. 
We insert 
\ie \label{tranz}
Z=i\frac{(1-Z')}{(1+Z')}
\fe 
into 
\eqref{ommdz}.
The meromorphic one-form $\omega_Z\textrm{d}Z$ then has the following form on the $Z'$-plane : 
\ie 
\omega_Z\textrm{d}Z&=-\frac{1}{2c}\frac{1}{(Z_++i)(Z_-+i)}   \frac{8 Z' \textrm{d}Z'}{(Z'-1)(Z'+1)(Z'-Z_+')(Z'-Z_-')},
\fe 
where 
\ie 
Z_+'=-\frac{i-Z_+}{-Z_+-i}, \quad
Z_-'=-\frac{i-Z_-}{-Z_--i}.
\fe 
Using the fact that $Z_+Z_-=-1$, we find that 
\ie 
Z_+'=-Z_-'.
\fe 

Now, under the map \eqref{tranz}, the boundary conditions \eqref{ybbc} are mapped to 
\ie \label{ybbc2}
(\tilde{R}+1)A |_1 &= (\tilde{R}-1)A |_{Z'_-}\\
(R+1)A |_{-1} &= (R-1)A |_{Z'_+}.
\fe 
Also note that when $c\rightarrow 0$, $Z_-' \rightarrow 1$ and $Z_+' \rightarrow -1$ .
As expected, the residues of $\omega$ at 1 and $Z_-'$ are equal and opposite, and the residues of $\omega$ at $-1$ and $Z_+'$ are equal and opposite.

Evaluating the formal gauge transformation of the gauge field at the poles of  $\omega$, we obtain 
\ie
&\left.A\right|_{1,Z_-'}=-\textrm{d} g g^{-1}+\left.\operatorname{Ad}_g \mathcal{L}\right|_{1,Z_-'},\left.\quad A\right|_{-1,Z_+'}=-\textrm{d} \tilde{g} \tilde{g}^{-1}+\left.\operatorname{Ad}_{\tilde{g}} \mathcal{L}\right|_{-1,Z_+'}
\fe
Since $\omega$ has zeroes at $0$ and $\infty$ shall use the following ansatz for the Lax operator : 
\ie 
\mathcal{L}=\left(B_{\xi}+ Z' J_{\xi}\right) \textrm{d} {\xi}+\left(B_{\bar{\xi}}+ (Z')^{-1} J_{\bar{\xi}}\right) \textrm{d} \bar{\xi}
\label{LBJ}
\fe 
where $B_{\xi,\bar{\xi}}$ and $J_{\xi,\bar{\xi}}$ are Lie-algebra-valued fields. 
Using this ansatz and the formal gauge transformation, we can find solutions for $J_{\xi}$ and $J_{\bar{\xi}}$, which are 
\ie 
J_{\xi} &= \frac{2 (j_{\xi}-\tilde{j}_{\xi})}{2(1+Z_-') + (1-Z_-')(R_g + \tilde{R}_{\tilde{g}}) },\\
J_{\bar{\xi}} &= \frac{2(j_{\bar{\xi}}-\tilde{j}_{\bar{\xi}})}{2(1+(Z_-')^{-1}) + (1-(Z_-')^{-1})(R_g + \tilde{R}_{\tilde{g}}) }
\fe 
where
\ie
j_\xi = g^{-1}\del_\xi g\quad,\quad 
\tilde{j}_\xi = \tilde{g}^{-1}\del_\xi \tilde{g}\;.
\fe
Using this expression in the Lax operator, and substituting the latter into the unifying action \eqref{grauni} obtained from 4d CS via formal gauge transformation, we obtain the action 
\ie \label{YB2}
S_{iYB}=\int_{{\mathbb{R}^2}}  (-i)\frac{Z_-}{c}\textrm{Tr} \left( (j_{\bar{\xi}} - \tilde{j}_{\bar{\xi}} )\frac{1}{1+\left(\frac{-iZ_-}{2} \right) (R_g +\tilde{R}_{\tilde{g}})}(j_{\xi} - \tilde{j}_{\xi} )\right)\txd\xi \wedge \txd \bar{\xi}.
\fe 
When $c\ra 0$ one has $Z_-/c \ra \rho$ and thus $Z_- \ra 0$, so that one retrieves the 
undeformed BM model \eqref{bmmodel} from this action in this limit. Thus for $c \neq 0$ 
this is a gravitational analogue of the inhomogeneous Yang--Baxter sigma model,
with target space $G/H$. 

In order to exhibit more clearly the coset structure of this sigma model, 
it is convenient to rewrite the sigma model action \eqref{YB2} using projection operators, 
\ie \label{projops}
P_{\mathfrak{h}}: \mathfrak{g} \rightarrow \mathfrak{h}, \quad P_{\mathfrak{m}}: \mathfrak{g} \rightarrow \mathfrak{m}, 
\fe 
where $\mathfrak{h}$ (already defined in \eqref{hdef}) 
and $\mathfrak{m}$ are the $\pm$ eigenspaces of $\eta$ respectively. 
Since the equivariance condition \eqref{equi} implies
\ie \label{equig}
\tilde{g}=\eta (g),
\fe 
 it follows that
\ie 
\left(R_g+\tilde{R}_{\tilde{g}}\right) \circ P_{\mathfrak{m}}({x})=2 P_{\mathfrak{m}} \circ R_g \circ P_{\mathfrak{m}}({x}),
\fe 
for ${x}\in \mathfrak{g}$, as shown, for example, in the appendix of \cite{Fukushima:2020dcp}.
This identity allows us to reexpress $B$ and $J$ in terms of projection operators and only the currents $j_{\xi}$
and $j_{\bar{\xi}}$, leading us to write the action in the form
\ie 
\label{YB2project}
S_{IYB}=\int_{{\mathbb{R}^2}}  (-i)\frac{Z_-}{c}\textrm{Tr} \left( j_{\bar{\xi}} P_{\mathfrak{m}}\left(\frac{1}{1-iZ_-  R_g \circ P_{\mathfrak{m}} }j_{\xi}\right)\right)\txd\xi \wedge \txd \bar{\xi}.
\fe 
Note that this action is real because (a) the volume 
element $(-i)
\txd\xi\wedge\txd\bar{\xi}$ is real, and (b) the kinetic term is real, because
$\overline{j_\xi} = j_{\bar{\xi}}$, and $iZ_- R_g$ is hermitian.

A characteristic feature of the above class of models 
is that the constant deformation parameter that usually appears
in front of $R_g$ in Yang--Baxter sigma models has been replaced
by $-iZ_-$. This can be understood as a worldsheet-dependent
deformation parameter. Moreover, the 
prefactor of the Lagrangian is also given by 
a factor of $-iZ_-/{c}$, which can now be thought
of as a worldsheet-dependent coupling.  This model thus falls into
the class of integrable gravitational Yang--Baxter sigma models
studied in \cite{Hoare:2020fye} and 
\cite{Cesaro:2025msv}, where it was
explained that proportionality of the worldsheet-dependent deformation
parameter and the coupling ensures the existence of a
flat Lax connection (something that is expected from our 4d
Chern--Simons perspective).

To obtain the explicit form for the Lax operator, we also need expressions for $B_{\xi}$ and $B_{\bar{\xi}}$, which were not needed in the derivations so far. These are 
\ie 
B_{\xi} &= \frac{j_{\xi} + \tilde{j}_{\xi}}{2} - \frac{1}{4} (R_g -\tilde{R}_{\tilde{g}})J_{\xi} (1-Z_-')\\
B_{\bar{\xi}} &= \frac{j_{\bar{\xi}} + \tilde{j}_{\bar{\xi}}}{2} - \frac{1}{4} (R_g -\tilde{R}_{\tilde{g}})J_{\bar{\xi}} (1-(Z_-')^{-1})
\fe 
The Lax operator then follows by substituting the explicit expressions for
$B_{\xi,\bar{\xi}}$ and $J_{\xi,\bar{\xi}}$ into \eqref{LBJ}.
By rewriting these explicit expressions in terms of the
projectors $P_{\mathfrak{h}}$ and $P_{\mathfrak{m}}$, the explicit form of the Lax 
operator can be shown to be 
\ie
\mathcal{L}_{\xi,\bar{\xi}}=P_{\mathfrak{h}}(j_{\xi,\bar{\xi}})-\frac{1}{2}P_{\mathfrak{h}}\circ R_g (\mp i Z_-)M_{\xi,\bar{\xi}} -i \left( \frac{Z-i}{-Z-i} \right)^{\pm 1} (\pm Z_- + i )M_{\xi,\bar{\xi}}
\fe 
where 
\ie 
M_{\xi,\bar{\xi}} =P_{\mathfrak{m}} \frac{1}{1\mp i Z_- R_g P_{\mathfrak{m}}}j_{\xi, \bar{\xi}}.
\fe 
This formula for the Lax operator is of a similar form to that found in \cite{Cesaro:2025msv}, but differs only in the factor $(\pm Z_- + i)$. This difference, however, is not surprising since we are working on a worldsheet with Euclidean signature, while Lorentzian signature was utilised in \cite{Cesaro:2025msv}. The flatness of this Lax operator follows from the third 4d CS equation of motion in \eqref{Leqs}.

\section{Gravitational Homogeneous Yang--Baxter Sigma Model}\label{Sec4}

In this section, we shall derive the homogeneous Yang--Baxter deformation of the BM model studied by Ces{\`a}ro and Osten \cite{Cesaro:2025msv} from 4d Chern--Simons theory. 

We shall utilise  the same 4d CS action used by Cole and Weck \cite{Cole:2024skp}, in the form given in \eqref{2ndactionB}, but with a different set of boundary conditions on the $Z$-plane. 
It shall be instructive to impose boundary conditions on fields defined on the $W$-plane
before proceeding to derive equivalent boundary conditions on the $Z$-plane. In the derivation of the BM model in \cite{Cole:2024skp}, the boundary condition at $W=\infty$ is the Dirichlet boundary condition $A_{\xi,\bar{\xi}}=0$.

More general boundary conditions will involve non-trivial relations
between $A$ and $\del_{u}A$ at $W=\infty$, where $\omega= -\textrm{d}u/u^2$, in terms of the coordinate $u=1/W$. 
In particular, we impose a more general mixed Dirichlet-Neumann boundary condition at $W=\infty$,
of the  form 
\ie \label{ubc}
A|_{u=0}=2\tilde{R}\partial_uA|_{u=0},
\fe 
where the factor of $2$ is merely a convenient choice. 
As in the derivation of the ordinary homogeneous Yang--Baxter sigma model, \cite{Delduc:2019whp},  it is sufficient that $\tilde{R}$ is a skew-symmetric matrix in order for the boundary equations of motion to be satisfied. 

Now, on the $Z$-plane, the boundary condition \eqref{ubc}
manifests itself as a pair of boundary conditions at the 
 poles of the meromorphic 1-form 
\ie \label{zoneform}
\omega_Z\textrm{d}Z =  \frac{{i}(\xi-\bar{\xi})}{4}\left(\frac{Z^2+1}{Z^2} \mathrm{~d} Z\right)\fe 
at
$Z=0$ and $Z=\infty$, which are 
\ie \label{bc3a}
A|_{Z=0}=\rho  R\left(\left.\partial_{Z} A\right|_{Z=0}\right),\quad  A|_{Z=\infty}=-  \rho \tilde{R}\left(\left.\partial_{\frac{1}{Z}} A\right|_{Z=\infty}\right).
\fe 
Here, by the equivariance condition \eqref{equi} the matrices $R$ and $\tilde{R}$ are related by 
\ie \label{rrtildeaut}
\tilde{R} \equiv \eta \circ R \circ \eta\; .
\fe 
The boundary conditions \eqref{bc3a} follow from \eqref{ubc} and \eqref{equi} since close to $Z=\infty$ we have $W\sim -\frac{\rho Z}{2}$, so that $\frac{\partial( 1/Z)}{\partial u} =-\frac{\rho}{2}$, which gives rise to the explicit $\rho$-dependence in the second boundary condition in \eqref{bc3a}, and likewise for the boundary condition at $Z=0$.
Crucially, these boundary conditions differ from those used to derive homogeneous Yang--Baxter sigma models in  \cite{Delduc:2019whp,Fukushima:2020dcp} since they involve explicit factors of the coordinate $\rho$.

The boundary conditions \eqref{bc3a} solve the boundary equations of motion, that take the general form 
\begin{equation}
\sum_{x=0,\infty}\left((\operatorname{res}_x \omega_Z\textrm{d}Z ) \epsilon^{i j}\textrm{Tr} ( A_i |_x \wedge \delta A_j |_x )+ \left(\operatorname{res}_x \xi_x \omega_Z\textrm{d}Z\right) \epsilon^{i j} \partial_{\xi_x}\textrm{Tr} (A_i \wedge \delta A_j ) |_x\right)=0 ,
\end{equation}
around the double poles of $\omega_Z \textrm{d}Z$,
where $\xi_x$ is the local coordinate in the vicinity of the location of a double pole, denoted $x$, and $i,j$ are summed over the coordinates $\xi$ and $\bar{\xi}$. In the present case, only the second term in the parenthesis has a nontrivial contribution. 
The boundary conditions \eqref{bc3a} solve the boundary equation of motion, since, e.g., the contribution to the boundary equation of motion at $Z=0$ is 
\ie \label{boundaryinnerproduct0}
(\textrm{res}_{0}Z\omega_Z\textrm{d}Z )\epsilon^{ij}\textrm{Tr}(\partial_{Z}A_i \wedge \delta A_j + A_i \wedge \partial_{Z}\delta A_j )
\fe
where $\textrm{res}_{0}Z\omega_Z\textrm{d}Z  = -\frac{\rho}{2}$, and using the first boundary condition, this is equivalent to 
\ie {\rho}\left(\textrm{res}_{0}Z\omega_Z\textrm{d}Z\right)\epsilon^{ij}\textrm{Tr}(\partial_{Z}A_i \wedge R \partial_{Z}\delta A_j + R \partial_{Z}A_i \wedge \partial_{Z}\delta A_j ),
\fe
which vanishes using the skew-symmetry of $R$. 
An analogous derivation works for the contribution to the boundary equation of motion at $Z=\infty$.

In order to obtain a more geometric
understanding of such boundary conditions, let us note first of all
that for any $G$ gauge field $A$, the pair consisting of $A$ and its holomorphic derivative ($\partial_WA$ or $\partial_ZA$ in the present example)  restricted to a point on $\mathbb{CP}^1$ transforms
like a $TG$-connection, with $TG$ denoting the tangent bundle group. The proof of this and other relevant facts and formulae regarding $TG$ are collected in Appendix \ref{tgapp}. The boundary conditions \eqref{bc3a} can be understood as restricting the gauge field and its holomorphic derivative to an isotropic subspace of $\mathfrak{tg}$, which is the Lie algebra of $TG$.
It may be helpful to draw an analogy with the derivation in the previous section, where the role of $TG$ was played by $G\times G$.

At this point, we have not imposed any other properties of the matrices $R$ and $\tilde{R}$.  To derive the IFT of interest, we shall further impose that $R$ and $\tilde{R}$ are solutions of the classical Yang--Baxter equation (cYBE); for example, $R$ satisfies 
\be
{} [R(x),R(y)] = R([R(x),y]-[R(y),x]), 
\ee
for $x,y\in \mathfrak{g}$.
The boundary conditions 
\eqref{bc3a} then correspond to restricting the gauge fields and their holomorphic derivatives to be valued in certain isotropic \textit{subalgebras} of $\mathfrak{tg}$, that is  
\ie \label{zrhobc}
\left(\left.A\right|_{Z=0},\left.\partial_{Z} A\right|_{Z=0}\right) \in \mathfrak{g}_{R}, \quad\left(\left.A\right|_{Z=\infty},\left.\partial_{1/Z} A\right|_{Z=\infty}\right) \in \mathfrak{g}_{-\tilde{R}},
    \fe 
    where 
    \ie \label{subalgrtg}
\lg_{R} &= \{ ( R(x), \frac{1}{\rho }x), x \in \lg\},\\
\lg_{-\tilde{R}} &= \{ (- \tilde{R}(\tilde{x}), \frac{1}{\rho }\tilde{x}), \tilde{x} \in \lg\}.
    \fe 
    In other words, we find that  
\ie 
& \tilde{R}\textrm{ solution of cYBE}\quad\leftrightarrow \quad \mathfrak{g}_{-\tilde{R}} \subset \mathfrak{tg}|_{\infty} , \, \\& {R}\textrm{ solution of cYBE}\quad\leftrightarrow \quad \mathfrak{g}_{{R}} \subset \mathfrak{tg}|_{0}.
\fe 
This is analogous to the fact that the boundary conditions used to derive the inhomogeneous Yang--Baxter deformation of the BM model correspond to subalgebras of $\mathfrak{g}\oplus \mathfrak{g}$, as summarised in \eqref{rsolinhom}. As in the previous section, due to $R$ and $\tilde{R}$ being related by the Lie algebra automorphism \eqref{rrtildeaut}, the resulting Lie algebras and their corresponding Lie groups are isomorphic. 
    
The subalgebra $\mathfrak{g}_{{R}}$ belongs to the family of isomorphic subalgebras 
\be \label{cdalg}
\lg_{R,\alpha,\beta} = \{ (\alpha R(x), \beta x), x \in \lg\}.
\ee
where $\alpha,\beta\in \mathbb{C}^*$, whose properties are reviewed in Appendix \ref{subscd}. 
In particular, the pair $(A,\partial_u A)$ that satisfies the boundary condition \eqref{ubc} is valued in the  $\mathfrak{g}_{\tilde{R},2,1}$ subalgebra  of $\mathfrak{tg}$, and is invariant under gauge transformations of the corresponding subgroup of $TG$.
This gauge invariance descends to the $Z$-plane, since the boundary conditions
\eqref{bc3a}
 are invariant under gauge transformations of the form \eqref{gaugetxtild} generated by the corresponding subgroups of $TG$. 
Moreover, the Dirichlet boundary conditions of \cite{Cole:2024skp} correspond to
the $\alpha \rightarrow 0$ limit of \eqref{cdalg}, whereby the nonabelian Lie
algebra $\mathfrak{g}_R$ reduces to the abelian algebra $\mathfrak{g}_{ab}$. In
this sense, the deformation of the BM model studied here corresponds to a 
deformation of the Abelian Lie algebra $\lg_{ab}$ to the non-Abelian Lie algebra
$\lg_R$. 

Choosing $R$ and $\tilde{R}$ to be solutions of the classical Yang--Baxter equation is crucial
for the integrability (in the sense of exhibiting a flat Lax connection) of the field theory we shall derive. Meanwhile, from the perspective of gauge theory, the classical Yang--Baxter equation implies that $\lg_{R} $ and $\lg_{-\tilde{R}}$ are subalgebras of $TG$, which ensures that the boundary conditions are gauge invariant with respect to their corresponding groups. 

As in the case of the inhomogeneous Yang--Baxter deformation derived in the previous section, the gauge invariance of the boundary conditions leads to degrees of freedom that are redundant and must be fixed. We shall also make use of the  formal gauge transformation \eqref{form} in this section.
In the present case, the boundary conditions we use involve holomorphic derivatives of the gauge field. Let us consider the boundary condition at $Z=0$. 
 The  holomorphic derivative of the gauge field at this point has the formal gauge transformation 
\ie 
\partial_Z (A)^{\hat{g}}|_{Z=0} = \hat{g}^{-1} (\partial_zA +\tilde{\textrm{d}}(\partial_Z\hat{g} \hat{g}^{-1})+[A,\partial_Z\hat{g}\hat{g}^{-1}])\hat{g}|_{Z=0},
\fe 
where $\hat{g}:\Sigma \times \mathbb{CP}^1 \rightarrow G$. 
A further ordinary gauge transformation with respect to $U:\Sigma \times \mathbb{CP}^1 \rightarrow G_R$ leads to 
\ie 
\partial_Z (A)^{\hat{g}U}|_{Z=0}= (\hat{g}U)^{-1} (\partial_zA +\tilde{\textrm{d}}(\partial_Z(\hat{g} U)(\hat{g}U)^{-1})|_{Z=0}+[A,\partial_Z(\hat{g}U)(\hat{g}U)^{-1}])\hat{g}U|_{Z=0}
\fe 
where
\ie 
\partial_Z(\hat{g} U)(\hat{g}U)^{-1}|_{Z=0}=\partial_Z\hat{g} \hat{g}^{-1}|_{Z=0}+\hat{g}\partial_ZU U^{-1}\hat{g}^{-1}|_{Z=0}.
\fe 
This is compatible with the group multiplication law of $TG$ defined in Appendix \ref{tgapp}
and the fact that $G_R$ is a subgroup of $TG$, whereby one finds 
\ie 
(\hat{g},\partial_Z\hat{g}\hat{g}^{-1}) (U,\partial_ZU U^{-1})=(\hat{g}U, \partial_Z\hat{g}\hat{g}^{-1}+\hat{g}\partial_ZUU^{-1}\hat{g}^{-1}).
\fe 
 By picking $U$ such that 
\ie 
\partial_ZUU^{-1}|_{Z=0}=-\hat{g}^{-1}\partial_Z\hat{g}|_{Z=0}
\fe 
the expression $\partial_Z(\hat{g} U)(\hat{g}U)^{-1}|_{Z=0}$ can be set to zero.
Hence, defining $\hat{g}'=\hat{g}U$, we find that $\partial_Z{\hat{g}'}|_{Z=0}=0$. In other words, we can use an ordinary $G_R$ gauge transformation to set to zero the edge mode corresponding to the restriction of the holomorphic derivative of $\hat{g}'$ to the pole of $\omega_Z\textrm{d}Z$ at $Z=0$. The equivariance condition \eqref{equi} further implies 
\ie \label{equiholg}
\partial_Z \hat{g}|_{Z=\infty}=\eta (\partial_Z \hat{g}|_{Z=0}),
\fe 
meaning that
the corresponding edge mode associated with $Z=\infty$ is set to zero as well.  
This gauge fixing of edge modes involving holomorphic derivatives is analogous to the gauge fixing in the previous section that led to the identification of pairs of edge modes in \eqref{pairfix}.

Let us now proceed to the derivation of the integrable field theory. 
We shall find it convenient to perform the coordinate transformation
\ie 
Z=i\frac{(1-Z')}{(1+Z')},
\fe 
which maps the double poles of $\omega$ at $Z=0,\infty$ to $Z' = \pm 1$.
Then, we find
\ie \label{rewrittenomega}
\omega_Z\textrm{d}Z = -i\rho \frac{Z'\textrm{d}Z'}{(1-Z')^2(1+Z')^2}.
\fe 

At the double poles of \eqref{rewrittenomega}, the boundary conditions take the form
\ie 
\left(\left.A\right|_{Z'=1},\left.\partial_{Z'} A\right|_{Z'=1}\right) \in \mathfrak{g}_{R,2i,\frac{1}{\rho}}, \quad\left(\left.A\right|_{Z'=-1},\left.\partial_{Z'} A\right|_{Z'=-1}\right) \in \mathfrak{g}_{\tilde{R},-2i,\frac{1}{\rho}},
    \fe 
in the notation introduced in \eqref{cdalg}.
To arrive at these expressions, we used $\frac{\partial(1/Z)}{\partial Z'}|_{Z'=-1} = 1/(i2)$.
 
The remaining derivation of the IFT is similar in many ways to the derivation of Yang--Baxter deformations of coset sigma models from 4d Chern--Simons theory given in \cite{Fukushima:2020dcp}, apart from crucial differences arising from $\rho$-dependence in various formulae. We perform a formal gauge transformation of the 4d CS gauge field, whereby we find 
\ie
&\left.A\right|_{Z'=1}=-\txd g g^{-1}+\left.\operatorname{Ad}_g \mathcal{L}\right|_{Z'=1},\left.\quad A\right|_{Z'=-1}=-\txd \tilde{g} \tilde{g}^{-1}+\left.\operatorname{Ad}_{\tilde{g}} \mathcal{L}\right|_{Z'=-1},
\fe
where $\hat{g}|_{Z'=1}=g$ and $\hat{g}|_{Z'=-1}=\tilde{g}$. To be precise, $\hat{g}$ here is in a gauge where its holomorphic detivatives at $Z'=1,-1$ are zero.

We will also postulate the following form for the Lax operator 
\begin{equation}
\label{hlax}
\mathcal{L}=\left(U_{\xi}+Z' V_{\xi}\right) \txd \xi+\left(U_{\bar{\xi}}+(Z')^{-1} V_{\bar{\xi}}\right) \txd \bar{\xi}
\end{equation}
that has the expected singular behaviour at the zeroes of $\omega_Z\textrm{d}Z$. 
Defining $j \equiv g^{-1} \textrm{d} g, \quad \tilde{j} \equiv \tilde{g}^{-1} \textrm{d} \tilde{g}$, and 
recalling that the holomorphic derivatives of $\hat{g}$ at $Z'=1,-1$ are gauge fixed to 0,
the boundary conditions can be shown to imply
\ie \label{jjtild}
&j_{ \xi}=U_{ \xi}+\left(1 - 2 \rho iR_g\right)\left(V_{ \xi}\right), \quad \tilde{j}_{ \xi}=U_{ \xi}-\left(1 - 2 \rho i\tilde{R}_{\tilde{g}}\right)\left(V_{ \xi}\right)\\&j_{ \bar{\xi}}=U_{ \bar{\xi}}+\left(1 + 2 \rho iR_g\right)\left(V_{ \bar{\xi}}\right), \quad \tilde{j}_{ \bar{\xi}}=U_{ \bar{\xi}}-\left(1 + 2 \rho i\tilde{R}_{\tilde{g}}\right)\left(V_{ \bar{\xi}}\right)
\fe 
where, e.g.,
\ie 
R_g \equiv \operatorname{Ad}_{g^{-1}} \circ R \circ \operatorname{Ad}_g.
\fe 

The expressions \eqref{jjtild} are equivalent to 
\ie 
\label{huv}
&U_{ \xi}=\frac{j_{ \xi}+\tilde{j}_{ \xi}}{2} +i\rho\left(R_g-\tilde{R}_{\tilde{g}}\right)\left(V_{ \xi}\right), \quad V_{ \xi}=\frac{1}{1 -i\rho R_g - i\rho \tilde{R}_{\tilde{g}}}\left(\frac{j_{ \xi}-\tilde{j}_{ \xi}}{2}\right)\\&U_{ \bar{\xi}}=\frac{j_{ \bar{\xi}}+\tilde{j}_{ \bar{\xi}}}{2} - i\rho\left(R_g-\tilde{R}_{\tilde{g}}\right)\left(V_{ \bar{\xi}}\right), \quad V_{ \bar{\xi}}=\frac{1}{1 + i\rho R_g +i\rho \tilde{R}_{\tilde{g}}}\left(\frac{j_{ \bar{\xi}}-\tilde{j}_{ \bar{\xi}}}{2}\right).
\fe 
We can then use the unifying action \eqref{grauni} to obtain 
\ie \label{YB1}
S_{hYB}=-i \int_{{\mathbb{R}^2}}\rho\textrm{Tr}\left(\frac{j_{{\xi}}-\tilde{j}_{{\xi}}}{2} \left(\frac{1}{1+i\rho R_g+i\rho\tilde{R}_{\tilde{g}}}\left(\frac{j_{\bar{\xi}}-\tilde{j}_{\bar{\xi}}}{2}\right)\right)\right) \txd\xi \wedge \txd \bar{\xi}.
\fe 
Just like the inhomogeneous Yang--Baxter deformation of the BM model, the coset
sigma model \eqref{YB1} with target space $G/H$ can be reexpressed using the projection operators \eqref{projops}, that is, it can be written as 
\ie \label{YB1project}
S_{hYB}=-i\int_{{\mathbb{R}^2}}\rho\textrm{Tr}\left({j_{\bar{\xi}}} P_{\mathfrak{m}}\left(\frac{1}{1+i2\rho R_g\circ P_{\mathfrak{m}}}j_{{\xi}}\right)\right) \txd\xi \wedge \txd \bar{\xi}.
\fe 
This action is real for the same reasons as those given after 
\eqref{YB2project} for the inhomogeneous case. Note that the 
worldsheet-dependent coupling is $\rho$, exactly as in the original
BM model \eqref{bmmodel}, but unlike that of the inhomogeneous model, where $\rho$ 
is replaced by the more general function $Z_-(\rho,z)$. 

By substituting the expressions \eqref{huv} for $U$ and $V$
into the Lax operator \eqref{hlax} and rewriting it
in terms of the
projectors $P_{\mathfrak{h}}$ and $P_{\mathfrak{m}}$, the 
Lax operator can be shown to have the explicit form 
\ie
\mathcal{L}_{\xi,\bar{\xi}}=P_{\mathfrak{h}}(j_{\xi,\bar{\xi}})\pm i \rho 2 P_{\mathfrak{h}}\circ R_g V_{\xi,\bar{\xi}} + \left( \frac{Z-i}{-Z-i} \right)^{\pm 1} V_{\xi,\bar{\xi}}
\fe 
where 
\ie 
V_{\xi,\bar{\xi}} =P_{\mathfrak{m}} \frac{1}{1\mp 2 \rho  R_g P_{\mathfrak{m}}}j_{\xi, \bar{\xi}}.
\fe 
This form of the Lax operator agrees with that presented in \cite{Cesaro:2025msv}.

We have thus derived the homogeneous Yang--Baxter deformation of
the BM model studied by Cesaro and Osten \cite{Cesaro:2025msv} from
4d Chern--Simons theory. The difference from the ordinary homogeneous
Yang--Baxter deformation of a coset sigma model is that the deformation
parameters appearing in front of $R_g$ and $\tilde{R}_{\tilde{g}}$
have now been replaced by the coordinate $\rho$ multiplied by $i$,
and the coupling is proportional to the coordinate $\rho$.  This
model is also a special case of the general Yang--Baxter sigma model
with worldsheet-dependent couplings studied by Hoare et al.
\cite{Hoare:2020fye}.


\appendix

\section{Derivatives on the \texorpdfstring{$W$}{W}-plane and the 
\texorpdfstring{$Z$}{Z}-plane}\label{derwz}

The derivatives along $\mathbb{R}^2$ in the basis where $W$ is fixed and the same derivatives in the basis where $Z$ is fixed are related as 
\ie \label{der2}
\partial_{\xi}|_W&=\partial_{\xi}|_Z+\left(\frac{\partial Z}{\partial \xi}\right)|_W\partial_Z,\\
\partial_{\bar{\xi}}|_W&=\partial_{\bar{\xi}}|_Z+\left(\frac{\partial Z}{\partial \bar{\xi}}\right)|_W\partial_Z,
\fe 
where $|_W$ and $|_Z$ indicate the variable being held fixed. In order to relate
these to the derivatives appearing in \eqref{shiftedders}, note that on the one
hand one has the standard identities
\be 
\left( \frac{\partial \xi}{\partial W}\right)|_Z=1/\left(\frac{\partial W}{\partial \xi}\right)|_Z
\quad,\quad
\left(\frac{\partial Z}{\partial \xi}\right)|_W \left(\frac{\partial \xi}{\partial W}\right)|_Z \left(\frac{\partial W}{\partial Z}\right)|_\xi=-1
\ee
and on the other hand from $\omega = \textrm{d}W$ one has 
\be
\omega = \textrm{d}W \quad\Ra\quad 
\omega_{Z} = \left(\frac{\del W}{\del {Z}}\right)|_{\xi} \quad,\quad
\omega_{\xi} = \left(\frac{\del W}{\del {\xi}}\right)|_{Z} 
\quad,\quad
\omega_{\bar\xi} = \left(\frac{\del W}{\del {\bar\xi}}\right)|_{Z} 
\ee
Thus, putting this together one finds  
\ie 
\left(\frac{\partial Z}{\partial \xi}\right)|_W =
-\frac{\omega_{{\xi}}}{\omega_Z} \quad,\quad
\left(\frac{\partial Z}{\partial \bar{\xi}}\right)|_W = -\frac{\omega_{\bar{\xi}}}{\omega_Z},
\fe 
meaning that the relations in \eqref{der2} agree precisely 
with those of \eqref{shiftedders} if we identify $\tilde{\partial}_{{\xi},{\bar{\xi}}}$ with ${\partial}_{{\xi},{\bar{\xi}}}|_W$  
and 
${\partial}_{{\xi},{\bar{\xi}}}$ with ${\partial}_{{\xi},{\bar{\xi}}}|_Z$.

\section{General Choices of \texorpdfstring{$\omega$}{omega}}\label{genomega}
In the work of 
Cole and Weck \cite{Cole:2024skp}, the 4d CS setup that realises the BM model is derived from 6d CS on twistor space. They show that, for some meromorphic 3-form $\Omega_0$ defined on twistor space $\mathbb{PT}$, 
\ie 
\iota_{X_{\phi}\wedge X_{\tau }}\Omega_0 =\textrm{d} W
\fe 
where $W$ is defined in \eqref{wdef}, where $\phi$ and $\tau$ can be identified with angular and time coordinates in 4d gravity. Notably, this means that 
\ie 
\iota_{X_{\phi}} \textrm{d}W&=0\\
\iota_{X_{\tau}} \textrm{d}W&=0.
\fe 

We would like to derive the most general function, denoted $V$, that satisfies equations of this form, i.e., 
\ie 
\iota_{X_{\phi}} \textrm{d}V&=0\\
\iota_{X_{\tau}} \textrm{d}V&=0,
\fe 
or equivalently
\ie \label{twoeq}
v^1 \frac{\partial V}{\partial v^1} + \zeta \frac{\partial V}{\partial \zeta}&=0\\
 \frac{\partial V}{\partial v^2} + \zeta \frac{\partial V}{\partial v^1 }&=0,
\fe 
where $v_1$, $v_2$ and $\zeta$ are complex coordinates on twistor space defined in \cite{Cole:2024skp}.

The first equation implies that $V$
is invariant under simultaneous rescalings of $v_1$ and $\zeta$, and is therefore a function of their ratio
\ie
V=F\left(\frac{v_1}{\zeta}\right).
 \fe
 Defining $\lambda=v_1/\zeta$, we have
 \ie
\frac{\partial V}{\partial  v_1}=\frac{\partial \lambda}{\partial v_1}\frac{\partial V}{\partial  \lambda}= \frac{1}{\zeta} \frac{\partial V}{\partial  \lambda}
 \fe

The second equation of \eqref{twoeq} is thus equivalent to
 \ie 
\frac{\partial V}{\partial \lambda} + \frac{\partial V}{\partial v_2}=0,
 \fe
 meaning that 
 \ie 
V=F\left( v_2- \frac{v_1}{\zeta}\right).
 \fe 
The combination $v_2-\frac{v_1}{\zeta}$ is identified with the coordinate $W$, as explained in \cite{Cole:2024skp}.

Thus, in general,
we can consider more general 1-forms instead of $\textrm{d}W$ for the 4d CS.
In general, we could have
\ie 
\omega = F (W) \textrm{d}W
\fe
where $F(W)$ is some function of $W$.

\section{The Lie Algebra \texorpdfstring{$\ltg$}{tg} of
\texorpdfstring{$TG$}{TG} and
\texorpdfstring{$TG$}{TG} Gauge Fields}\label{tgapp}
In what follows, we collect
some of the relevant facts and formulae regarding the tangent bundle group
$TG$ of a Lie group $G$:
\begin{itemize}
\item In a right-invariant trivialisation, 
the group $TG$ can be identified with the semi-direct product of 
$G$ with its Lie algebra $\lg$, 
\be
\label{TGsdp}
TG \simeq G \times_{\textrm{Ad}} \lg
\ee
Thus the group
$TG$ consist of pairs $(g,v)\in G \times \lg$,
with the semi-direct product multiplication law
\be
\label{tgm}
TG \simeq \{(g,v)\in G \times\lg:\quad (g,v)(h,w) = (gh, v + \textrm{ad}_g w)\}\;\;.
\ee
\item The Lie algebra of $TG$, denoted $\mathfrak{tg}$, is the 
semi-direct sum of $\lg$ with $\lg_{ab}$,
\be
\label{adtg}
\mathsf{Lie}(TG) \equiv \mathfrak{tg} \simeq \lg \oplus_{\textrm{ad}}\lg_{ab}
\ee
with commutator 
\be
\label{tgc}
[(x,v),(y,w)] = ([x,y],[x,w]-[y,v]).
\ee
\item
For any semi-simple Lie algebra $\lg$, with non-degenerate
scalar product given by the 
trace $\Tr$, say, the Lie algebra $\mathfrak{tg}$ possesses 
an invariant scalar product given by 
\be
\label{tgsp}
\langle (x,v),(y,w)\rangle  = \Tr(xw + yv).
\ee
\item With respect to this scalar product, the Lie algebras
$\lg$ and $\lg_{ab}$ are both isotropic, 
\be
\langle (x,0),(y,0)\rangle  = 0 \quad,\quad \langle (0,v),(0,w)\rangle  = 0.
\ee
Thus \eqref{adtg} gives a semi-direct sum decomposition of $\mathfrak{tg}$ 
into two isotropic subalgebras $\lg$ and $\lg_{ab}$. 
\end{itemize}

We now look at $TG$ gauge fields and their gauge transformations:
\begin{itemize}
\item 
A $TG$ gauge field is a pair of 1-forms $(A,B)$ taking values in
$\mathfrak{tg}$, transforming under $TG$ gauge transformations 
as
\be
\label{tggt}
(A,B) \mapsto (A,B)^{(g,v)} = 
(A^g \equiv g^{-1} Ag + g^{-1} \textrm{d}g ,g^{-1} (B+ \textrm{d}_Av) g )
\ee
where
\be
\textrm{d}_Av = \textrm{d}v + [A,v] \;\;.
\ee
\item
The infinitesimal version of \eqref{tggt} is
\be
\label{itggt}
\textrm{d}_{(x,v)}(A,B)= \textrm{d}_{(A,B)} (x,v) = (\textrm{d}_Ax, [B,x] + \textrm{d}_Av)
\ee 
since (by the $\mathfrak{tg}$ commutation relations)
\be
\begin{aligned}
\textrm{d}_{(A,B)}(x,v) &= (\textrm{d}x,\textrm{d}v) + [(A,B),(x,v)] \\
&= (\textrm{d}x,\textrm{d}v) + ([A,x],[A,v] + [B,x])\\
&= (\textrm{d}_Ax,\textrm{d}_Av + [B,x]).
\end{aligned}
\ee
\end{itemize}

We can now establish the claim made in the main text that the pair consisting of the restrictions of a gauge field and its holomorphic derivative to a point on $\mathbb{CP}^1$
 transform like a $TG$-connection. For example, consider the pair $(A,\del_{Z}A)|_{Z}$.
Indeed, if $A \ra A^g$ one finds 
\be
\begin{aligned}
\del_{Z} A \ra \del_{Z}(A^g) &= g^{-1}\del_{Z}A g + (\del_{Z} g^{-1})A g + g^{-1} A
\del_{Z} g + \del_{Z} (g^{-1}\textrm{d}g) \\
&=g^{-1} \left( 
\del_{Z} A + \textrm{d}_A ((\del_{Z} g) g^{-1})\right) g .  
\end{aligned}
\ee
At fixed $Z$ this is precisely a $TG$ boundary gauge transformation \eqref{tggt}, with 
$g\ra g|_{Z}$ and $v \ra (\del_{Z} g) g^{-1}|_{{Z}}$.

\section{Subalgebras \texorpdfstring{$\mathfrak{g}_{R}\subset \ltg$}{g(R) of g}
and Solutions of the
cYBE}\label{subscd}


Consider the subspace 
\ie \label{grdeff}
\mathfrak{g}_R \equiv (R(x),x)
\fe 
of $\mathfrak{tg}$, where $R$ is an anti-symmetric matrix and $x\in \mathfrak{g}.$
It follows that $\lg_R$ is an isotropic 
subspace of $\mathfrak{tg}$ with respect to the scalar product  \eqref{tgsp}
since $R$ is anti-symmetric. Indeed, the restriction of $\langle .,.\rangle$ to 
$\lg_R$ is 
\be
\langle (R(x),x),(R(y),y)\rangle  = \Tr R(x)y + \Tr xR(y)  = \Tr R(x)y - \Tr R(x)y  = 0. 
\ee

The set $\lg_R$ of pairs $(R(x),x)$ 
is actually a subalgebra of the Lie algebra $\mathfrak{tg}$, with the commutator relations
given in \eqref{tgc}, precisely when $R$ satisfies the classical Yang--Baxter equation, 
\be
\lg_R \subset \mathfrak{tg} \quad\leftrightarrow \quad [R(x),R(y)] = R([R(x),y]-[R(y),x]). 
\ee
Thus, taken together, the two conditions on $R$ are equivalent to the statement
that $\lg_R$ is an isotropic Lie subalgebra of $\mathfrak{tg}$. 
For any such $R$ one has a vector space direct sum decomposition of 
the Lie algebra $\mathfrak{tg}$ into the Lie algebra 
$\lg$, identified with the set of pairs $(x,0)$, 
and the Lie algebra $\lg_R$, as can be seen by writing 
\be
(x,v) = (x-R(v),0) + (R(v),v) \in \lg \oplus \lg_R.
\ee
Since, as noted before, $\lg$ is also an isotropic subalgebra of $\mathfrak{tg}$, 
this provides a vector space decomposition of $\mathfrak{tg}$, into two isotropic
subalgebras. 
However, unlike the decomposition \eqref{adtg} this is 
\textit{not} a semi-direct sum decomposition since 
\be
{}[(x,0),(R(w),w)] = ([x,R(w)], [x,w]) \in \lg \oplus \lg_R .
\ee
Thus, unlike \eqref{TGsdp}, the group $TG$ cannot be regarded as 
a semi-direct product of $G$ and a group $G_R$ (with Lie algebra
$\lg_R$) 
(but is some more general ``twisted product'' of $G$ and $G_R$). 

Now, crucial to our application is the fact that the Lie algebra $\lg_R$ is largely independent of the choice of 
parameters $\alpha,\beta\in \mathbb{C}$ in 
\be
\lg_{R,\alpha,\beta} = \{ (\alpha R(x), \beta x), x \in \lg\}.
\ee
Indeed for $\alpha \beta\neq 0$ all these algebras are isomorphic. 
On the other hand, 
for $\alpha=0$ respectively $\beta=0$ one has 
\be
\lg_{R,0,\beta} \cong \lg_{\mathrm{ab}} \quad,\quad
\lg_{R,\alpha,0} \cong \lg.
\ee
In particular, in the limit $\alpha \ra 0$ the group $G_R$ In\"on\"u-Wigner
contracts to the Abelian group 
\be
\alpha \ra 0 \quad\Rightarrow\quad G_R \ra G_{\mathrm{ab}} = \lg_{\mathrm{ab}}.
\ee

Also, for any $\mathbb{Z}_2$ automorphism, $\eta$, of $\mathfrak{g}$, we have isomorphic Lie algebras $\mathfrak{g}$ and $\eta\mathfrak{g}$ and the matrix 
 \ie 
\tilde{R} : =\eta \circ R \circ \eta 
 \fe 
 then defines a Lie algebra $\mathfrak{g}_{\tilde{R}}$, following \eqref{grdeff},
which is isomorphic to $\mathfrak{g}_{{R}}$.

\bibliographystyle{ytphys} 
\bibliography{integravitability2}

\end{document}